\documentclass[5p,singleauthorgroup]{elsarticle}
\usepackage{xcolor}
\usepackage{lineno,hyperref}
\usepackage{epstopdf}
\usepackage{booktabs}
\modulolinenumbers[5]
\begin{document}
\renewcommand{\thefootnote}{\fnsymbol{footnote}}

\journal{Astroparticle Physics}











\begin{frontmatter}

	\title{Observations of the Crab Nebula with MACE (Major Atmospheric Cherenkov Experiment)}
\cortext[cor]{Corresponding author}
\author[add1]{Borwankar C.\corref{cor}}
\ead{chinmay@barc.gov.in}
\author[add1]{Sharma M.\corref{cor}}
\ead{mradul@barc.gov.in}
\author[add1]{Hariharan J.}
\author[add1]{Venugopal K.}
\author[add1]{Godambe S.}
\author[add1]{Mankuzhyil N.}
\author[add1]{Chandra P.\corref{cor}}
\ead{chandrap@barc.gov.in}
\author[add1,add2]{Khurana M.}
\author[add1]{Pathania A.}
\author[add1]{Chouhan N.}
\author[add1,add2]{Dhar V. K.}
\author[add1]{Thubstan R.}
\author[add1]{Norlha S.}
\author[add1]{Keshavanand}
\author[add1,add2]{Sarkar D.}
\author[add1]{Dar Z. A.}
\author[add1]{Kotwal S. V.}
\author[add1]{Godiyal S.}
\author[add1]{Kushwaha C. P.}
\author[add1,add2]{Singh K. K.}
\author[add1]{Das M. P.}
\author[add1,add2]{Tolamati A.}
\author[add1,add2]{Ghosal B.}
\author[add1]{Chanchalani K.}
\author[add1]{Pandey P.}
\author[add1]{Bhatt N.}
\author[add1,add2]{Bhattcharyya S.}
\author[add1,add2]{Sahayanathan S.}
\author[add1]{Koul M. K.}
\author[add3]{Dorjey P.}
\author[add3]{Dorji N.}
\author[add3]{Chitnis V. R.}
\author[add1]{Tickoo A. K.\fnref{fn1}}
\author[add1,add2]{Rannot R. C.}
\author[add1,add2]{Yadav K. K.}

\address[add1]{Astrophysical Sciences Division, Bhabha Atomic Research Centre, Mumbai 400085, India.}
\address[add2]{Homi Bhabha National Institute, Mumbai 400094, India.}
\address[add3]{Department of High Energy Physics, Tata Institute of Fundamental Research, Mumbai, India.}

\fntext[fn1]{Deceased}




\begin{abstract}
		The Major Atmospheric Cherenkov Experiment (MACE) is a 
		large size (21m) Imaging Atmospheric Cherenkov Telescope (IACT) installed 
		at an altitude of 4270m above sea level at Hanle, Ladakh 
		in northern India. Here we report the detection of Very High 
		Energy (VHE) $\gamma$-ray emission from Crab Nebula above 80 GeV. 
		We analysed $\sim$ 15 hours of data collected at low zenith angle 
		between November 2022 and February 2023. 
		The energy spectrum is well described by a log-parabola function 
		with a flux of $\sim$ ($3.46 \pm 0.26_{stat}) \times 10^{-10}$ 
		$\textrm TeV^{-1} \textrm cm^{-2} \textrm s^{-1}$, 
		at 400 GeV with spectral index of $2.09 \pm 0.06_{stat}$ and a 
		curvature parameter of $0.08 \pm 0.07_{stat}$. The $\gamma$-rays 
		are detected in an energy range spanning from 80\,GeV 
		to $\sim$5\,TeV. The energy resolution improves from $\sim$ 
		34\% at an analysis energy threshold of 80\, GeV to 
		$\sim$ 21\% above 1\,TeV. The daily light curve and the 
		spectral energy distribution obtained for the Crab Nebula 
		is in agreement with previous measurements, considering 
		statistical and systematic uncertainties.
\end{abstract}

\begin{keyword}
gamma rays: general -- telescopes -- techniques: miscellaneous -- methods: data analysis -- individual: Crab Nebula
\end{keyword}

\end{frontmatter}


\section{Introduction}
\label{sec:intro}
			The Crab Nebula, a supernova remnant (SNR), located at a distance of 
			$\sim$ 2 kpc \citep{crabDistance}, was detected in 1054 A.D. 
			\citep{carb1054II,crab1054}. It remains one of the most widely studied 
			astrophysical sources across various wavelengths of electromagnetic spectrum 
			\citep{crabRadio,crabIR,crabOptical,crabUV,crabXrays,rannot2002,CrabMagic,tickoo2014,CrabHEGRA,crabAstriHorn,CrabMAGIChighZ}. 
			Observations of emission from the nebula region at every accessible 
			wavelength have resulted in a well-determined broadband spectral 
			energy distribution (SED) of the Crab Nebula. 
			

			In a widely accepted classical model of the Crab Nebula, it consists 
			of a pulsar at its center, a shocked pulsar wind nebula, a bright 
			expanding shell of thermal gas or thermal filament, and a faint 
			freely expanding supernova remnant located outside the edge of nebula 
			\citep{Davidson1985,crabReview}. The pulsar injects energy in the 
			form of a magnetized, cold, and ultra-relativistic wind of 
			electron-positron pairs at a constant rate into the nebula 
			region. The randomized electron-positron pairs produce 
			synchrotron radiation from radio to high energy (HE) 
			$\gamma$-rays of few 100 MeV. 
			The upscattering of low energy 
			photons by the relativistic electron-positron pairs via inverse 
			Compton (IC) process leads to the emission of very high energy 
			(VHE) $\gamma$-ray photons in the range from few GeV to 100 TeV 
			from the Crab nebula \citep{AtoyanAharonian,Meyer2010,HAWC}. These 
			underlying radiative processes predict time-independent 
			luminosity of the source over a timescale of few hundred 
			years \citep{Buhler2014}. However, several rapid and bright 
			flares of different magnitudes have been observed from the 
			Crab Nebula since 2009 by the space-based $\gamma$-ray 
			detectors in the HE $\gamma$-ray band 
			\citep{Abdo2011,Tavani2011,Buehler2012,Mayer2013,Striani2013,Arakawa2020}. 
			The measured enhancements in HE fluxes up to few GeV are attributed to the 
			synchrotron component of the nebula emission while VHE fluxes from the 
			IC component have shown no significant changes and remain consistent 
			with the average emission level \citep{Aliu2014,HESS2014}. Thus, Crab Nebula 
			remains a constant and steady emitter of VHE photons and has been used as a 
			standard candle \citep{crabStandardCandle2021} in the field of 
			ground-based gamma-ray astronomy for more than three decades. 
			It is also used for cross-calibration of telescopes, monitoring 
			of the instrument response over time, and expressing the emission 
			level of other astrophysical sources in the Crab Unit (C.U.) in 
			observational astronomy.

%
%
			
			The development of Imaging Atmospheric Cherenkov Telescopes (IACTs) 
			for ground-based VHE astronomy has unveiled the $\gamma$-ray sky 
			in an unprecedented way. Today, more than 270 \footnote{http://tevcat2.uchicago.edu/} sources of GeV-TeV 
			$\gamma$-rays have been discovered in the known universe over a 
			short time span with IACTs being a major actor in the field 
			\citep{crab1989,Sitarek2022,Almeida2022}.
			With the aim of detecting $\gamma$-rays of energies down to $\sim$ 
			30 GeV with high point source sensitivity, a state-of-the-art IACT, 
			MACE (Major Atmospheric Cherenkov Experiment) setup at Hanle, 
			Ladakh in northern India, has been 
			carrying out regular observations of Crab Nebula since September 
			2021. In this paper, we present the first light results from 
			Crab Nebula using MACE observations conducted between 
			November 2022 to February 2023, totalling $\sim$15 hours 
			of observation time. We estimate the differential energy 
			spectrum of Crab Nebula in the zenith angle range of 
			$10^\circ$ to $35^\circ$. The analysis covers both the ON 
			source and the OFF source region during the observation period. 


			This paper is organized as follows: Section $~\ref{sec:MACE}$ 
			provides an introduction to MACE. Section $~\ref{sec:MCS}$ 
			details about the Monte Carlo (MC) simulations carried out for MACE. Section 
			$\ref{sec:analysis}$ covers the Observations and Data analysis
			practices employed in handling the simulation and observation data. 
			The validation of the MC simulations by comparison with the observed 
			data is discussed in Section $~\ref{sec:comparison}$. The performance 
			of MACE telescope is evaluated by estimating various parameters 
			like Energy threshold, Energy and angular resolution and integral 
			flux sensitivity in Section $~\ref{sec:perform}$. Finally we conclude 
			by discussing the results obtained and conclusion in Section $~\ref{sec:results}$.

\section{The MACE Telescope}
\label{sec:MACE}
		MACE is the latest major step in the long-development of 
		ground-based VHE gamma-ray astronomy program in India \citep{Singh2021}.
		It indirectly detects VHE $\gamma$-ray photons emitted from the 
		galactic or extragalactic sources by collecting the Cherenkov 
		light produced by the cascade of relativistic charge particles 
		in an extensive air shower, initiated by the interaction of GeV-TeV 
		$\gamma$--rays, within the Earth-atmosphere \citep{Ong1998}.
		MACE is located at Hanle 
		(32$^\circ$ 46$^{'}$ 46$^{''}$ N, 78$^\circ$ 58$^{'}$ 35$^{''}$ E) 
		in the Himalayan ranges of North India \citep{mace1,mace2,mace3}. 
		Located at an altitude of 4270m above sea level, it stands 
		as the only operational IACT such a high altitude. The telescope has 
		a 21m diameter quasi-parabolic reflector with an f-ratio of 1.2 
		and a focal length of 25m. Its light collector consists of 356 
		mirror panels, each measuring 984 mm $\times$ 984 mm. 
		Each panel is composed of four 488 mm $\times$ 488 mm facets 
		of spherical mirrors made of aluminum with SiO$_2$ coating.
		The total light collector area amounts to $\sim 337 m^{2}$. 

\begin{figure}[!h]
\begin{center}
\includegraphics[width=0.40\textwidth, angle =0, clip]{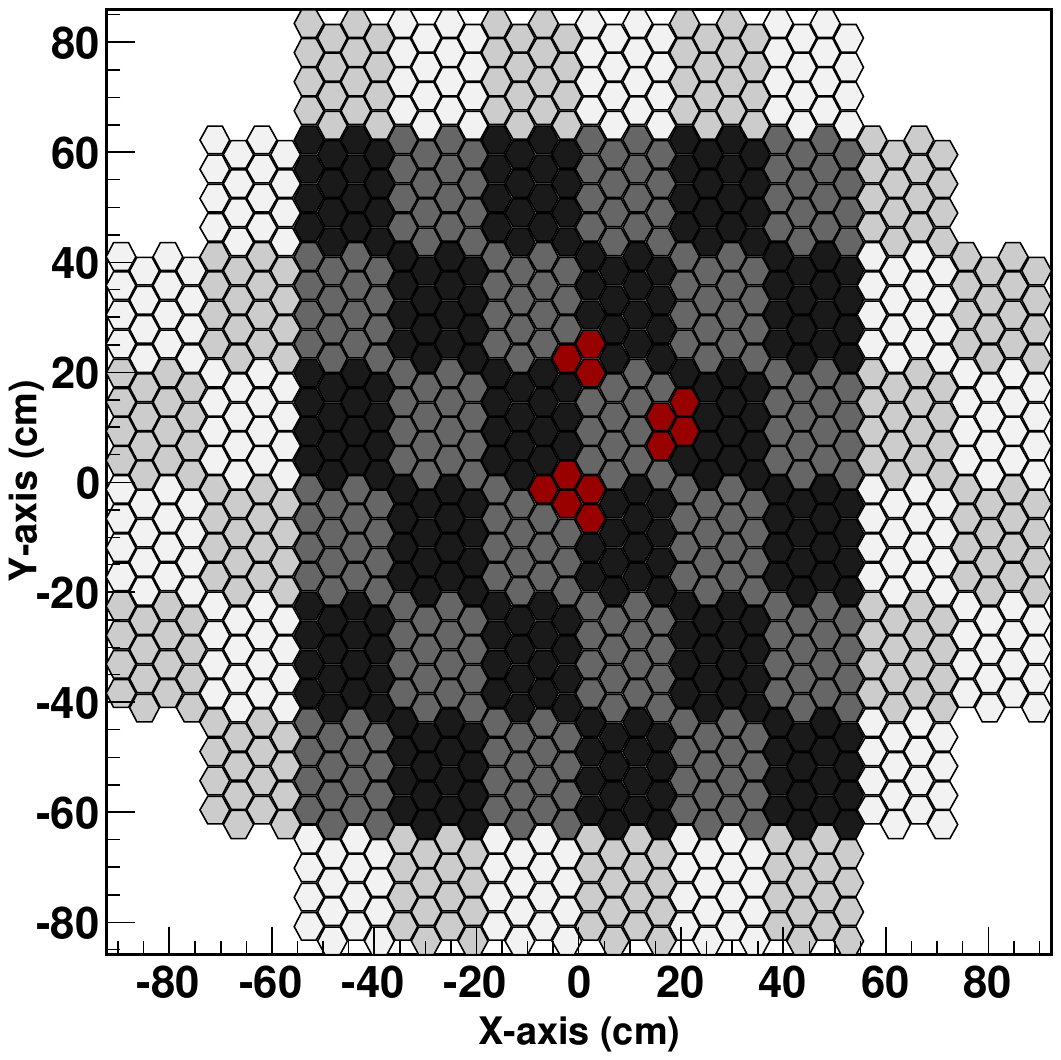}
\caption[The MACE camera layout]{The MACE camera configuration consists of 1088 
	pixels organized into 68 modules, each containing 16 PMTs (photo-multiplier tubes) \citep{mace2}. 
	The trigger signal originates from a darker shaded region encompassing 576 pixels. 
	The typical trigger patterns for different programmable schemes (3, 4, 
	5 CNNN :-Close Cluster Nearest Neighbor) are depicted in red. The 
	trigger configuration of 4 (CCNN) pixels is implemented in MACE.}\label{Figure:maceCam}
\end{center}
\end{figure}

		The mirror facets have a graded focal length ranging from 
		25-26.52m \citep{vkd2022}. The graded focal length ensures 
		that the on-axis spot size is minimum at the focal plane. 
		The MACE employs the Active Mirror Control system to align 
		individual mirrors panels through computer-controlled actuators 
		to ensure optimal focusing at all zenith angles. The MACE 
		camera \citep{maceCamera} is equipped with 1088 photomultiplier 
		tubes (PMTs). The entire camera is segmented into 68 sections, 
		referred to as Camera Integrated Modules (CIM). Each CIM 
		consists of 16 PMTs with a uniform pixel resolution of $0.125^{\circ}$ and 
		fitted with a Compound Parabolic Concentrator (CPC) featuring a 
		hexagonal aperture of size 0.125$^\circ$ and a circular exit 
		that matches the active photo-collection area of the PMT. 
		CPCs eliminate the dead space between PMTs. 
		Out of a total of 68 CIMs, the innermost 36 CIMs consisting 
		of 576 PMTs are utilized for MACE trigger generation. 
		The field of the view for the total trigger region measures 
		$\sim 2.62^{\circ} \times 3.02^{\circ}$ whereas the optical 
		field of view of MACE camera is $\sim 4.36^{\circ} \times 4.03^{\circ}$. 
		The MACE camera layout with 3, 4 and 5 CCNN trigger patterns is shown in 
		Figure \ref{Figure:maceCam}.

\section{Monte Carlo simulations}
\label{sec:MCS}
		One of the main challenges in the IACT data analysis 
		arises from the absence of calibrated sources of TeV 
		$\gamma$-rays. Monte Carlo simulations offers a solution 
		by simulating the expected Cherenkov radiation produced 
		from $\gamma$-rays and cosmic rays induced showers and 
		modelling the response of an IACT to this light. 
		A comprehensive database of simulated images of $\gamma$ and 
		hadron-induced air showers provides a necessary calibration to 
		reconstruct particle type, energy, and arrival direction of the 
		events detected by an IACT. This includes the development of 
		Machine Learning (ML) models for estimating the type and energy 
		of the events, as well as calculation of instrument response 
		functions of the telescope.

		The development of $\gamma$ and cosmic ray induced air showers through 
		the atmosphere and the production of Cherenkov light were simulated using 
		CORSIKAv6.990 \citep{CORSIKA}. The atmospheric
		density profile and extinction coefficients for visible light are estimated
		using U.S. standard atmospheric model. We use the IACT-Bernlohr routines \citep{Bern} 
		from CORSIKA to model the propagation and absorption of Cherenkov radiation 
		through the atmosphere up to the ground level. These routines provide
		data about the position, orientation, wavelength, time of arrival and
		number of Cherenkov photons emitted during EAS development. The Cherenkov
		photons which intersect with the fiducial sphere of the size equal to
		telescope radius located at telescope position are stored.

		We simulated a total of 24.8 $\times$ 10$^{6}$ 
		$\gamma$-ray induced showers and 4.96 $\times$ 10$^{8}$ proton induced
		air	showers respectively. The $\gamma$-ray showers were divided into 
		three independent samples with sizes of 8.9 $\times$ 10$^{6}$, 2.5 $\times$ 10$^{6}$ 
		and 13.4 $\times$ 10$^{6}$ events respectively. These samples are used
		for training random forest models for classification and regression,
		for validation of trained random forest models and for estimation
		of various telescope response matrices respectively. We estimate
		background rates in MACE data using dedicated OFF runs (see section \ref{sec:analysis})
		and hence we do not require evaluation of response matrices for cosmic rays.
		The simulated proton induced air showers were thus divided into 2 
		independent samples consisiting of 3.96 $\times$ 10$^{8}$ and 1 $\times$ 10$^{8}$
		events used for random forest training and validation.
		A small set of Helium induced air showers consisting of few million showers
		in the zenith range of 15$^{\circ}$ - 25$^{\circ}$ is also generated using CORSIKA.
		The Helium showers are used only for the comparison of observed event rates
		with integral rates	estimated from the simulations(see section \ref{sec:comparison}).
		The number of events used for training, validation and generation of
		response matrices are summarised in Table \ref{simtab}

		\begin{table}[]
			\tabcolsep=0.10cm
		\begin{tabular}{|l|ll|ll|}
		\hline
		& \multicolumn{2}{c|}{$\gamma$} & \multicolumn{2}{c|}{ Proton} \\ \hline
		& \multicolumn{1}{l|}{All} & Triggered & \multicolumn{1}{l|}{All} & Triggered \\ \hline
		Training & \multicolumn{1}{l|}{8.9 $\times$ 10$^{6}$} & 60000 & \multicolumn{1}{l|}{3.96 $\times$ 10$^{8}$} & 60000 \\ \hline
		Validation & \multicolumn{1}{l|}{2.5 $\times$ 10$^{6}$} & 15000 & \multicolumn{1}{l|}{1.0 $\times$ 10$^{8}$} & 15000 \\ \hline
		Response & \multicolumn{1}{l|}{13.4 $\times$ 10$^{6}$} & 90000 & \multicolumn{1}{l|}{-} & - \\ \hline
		\end{tabular}
		\caption{\label{simtab} Number of events used for training validation and 
			estimation of telescope response functions}
		\end{table}

		
		The $\gamma$-rays were simulated with a differential energy spectrum 
		given by dN/dE $\propto$ $E^{-2.59}$ in the energy range of 10 GeV to 
		20 TeV whereas the cosmic ray protons were simulated in the energy 
		range of 20 GeV to 20 TeV with a spectral index of 2.7. All the showers 
		were generated for zenith angles ranging from $10^\circ$ to $35^\circ$ 
		in increments of $1^\circ$, to cover the zenith range of Crab data 
		analysed in this work. The azimuth dependence of extensive air showers 
		was incorporated by simulating showers at azimuth angles ranging from 
		$0^\circ$ to $315^\circ$ in increments of $45^\circ$ for each zenith. 
		To accelerate simulations and address low global trigger efficiency, 
		each generated shower was utilized 10 times, with the telescope placed 
		at 10 randomized locations within a 650 m circle around the shower axis. 
		The protons were simulated with their arrival direction within 
		3$^{\circ}$ of the telescope pointing direction. The Crab nebula being
		the point source for IACTs, we keep arrival direction of all simulated
		$\gamma$-rays along the telescope pointing direction.

		The quantum efficiency of MACE PMTs and reflectivity of the 
		MACE mirrors is accounted for by using the CEFFIC option of the
		Bernlohr routines. The wavelength dependence of quantum efficiency
		and reflectivity of MACE telescope is reported in \citep{CBTrigger}.
		We developed a C++/ROOT based program to simulate the response of 
		MACE \citep{MACESIM} to air showers. The program simulates the reflector, the 
		camera geometry and the camera electronics. While specular reflection is 
		used to ray trace individual Cherenkov photons from the mirror surfaces, 
		the finite optical Point Spread Function (PSF) of the reflector is
		simulated using Gaussian smearing of reflected direction. The smearing
		angle is set to a value such that the simulated optical PSF matches
		with the optical PSF measured during mirror alignment observations
		performed on the pole star. Single photo-electron pulses are added to each PMT 
		with an average rate of 293 MHz to account for the contribution from light of 
		night sky (LONS). The rate of LONS induced photo-electrons is estimated 
		from the observations reported by nearby Himalayan Chandra Telescope (HCT) 
		\citep{Stalin,CBTrigger}. The detailed simulations of
		various parts of camera electronics related to trigger and data acquisition are
		performed and shower images which pass preset trigger criterion are stored
		in ROOT files. To realistically simulate the electronics the PMT
		pulses are modeled by superposition of the single photo-electron
		pulses, where each single photo-electron pulse is approximated by
		a Gaussian shape. The amplitude of the single photo-electron pulse
		is set on the basis of lab-measured gain of reference PMTs.
		The MACE event trigger is generated through two level trigger system,
		where a first level trigger checks the coincidence of threshold crossing
		pulses within a module of 16 pixels (see Figure \ref{Figure:maceCam}).
		The second level trigger detects the events which are spread over 
		pixels located in 2 or more than 2 neighbouring modules. The coincidence
		gate width of first level trigger is 5 ns, where as it is 10 ns 
		for second level. The MACE simulation program implements the two
		level trigger system including the pulse shaping at outputs of PMT
		pulse discriminators, time coincidence of output square pulses generated by each PMT
		as well as first level trigger (FLT) modules. The amplification of
		PMT pulses by low and high gain channels, digitisation of PMT pulses
		by DRS4	12 bit- analog to digital converter (ADC) assembly, various
		jitters in trigger and data acquisition
		are simulated to generate the output images in a format equivalent
		to the real MACE data. The output images in simulations thus consist
		of 2 arrays with length equal to number of PMTs, consisting data of
		of integrated digital counts for high and low gain pathways. The output images
		also have pulse profile data and information on ADC saturation status
		for high and low gain paths. The complete analysis chain described
		in Section \ref{sec:analysis} is then applied on the simulated images.


		The telescope response simulation, used to estimate various performance 
		parameters as described in earlier studies \citep{mace2,mace3} 
		did not include the camera electronics such as trigger and data 
		acquisition system. The old trigger simulation consisted
		of a topological trigger where single channel discrimination threshold
		was simulated in terms of number of photo-electrons collected at
		a PMT. The simulated image consisted of an array of number of
		photo-electrons collected at each PMT.
		All machine learning models and instrument response functions used
		in the analysis of Crab Nebula observations have been generated
		using the new simulation program that includes simulation of complete
		electronics chain.

		The Crab field of view contains 3 bright stars namely, $\zeta$-Tauri 
		(Mag 2.91), $\omega$-Tauri (Mag 4.81) and 121-Tauri (Mag 5.34) at 
		an angular distance of 1.2$^{\circ}$, 1.5$^\circ$ and 1.7$^{\circ}$ 
		from the center of FoV of Crab, respectively. During the Crab 
		observation runs, MACE PMTs were operated at a low gain 
		of $\sim$ 24000, corresponding to an average single 
		photo-electron pulse height of 8 \emph{mV}. However, 
		the presence of bright stars and high average ambient LONS
		in galactic plane, disabled some PMTs due to high anode
		current and trigger masking of some channels occurred near the position
		of the bright stars. To minimize the number of disabled and trigger
		masked pixels, we operated MACE with a high discrimination
		threshold of 90 \emph{mV} during Crab observations. The operational
		configuration corresponds to single channel discrimination threshold
		of $\sim$ 11 photo-electrons with a 4 close cluster near neighbor trigger 
		(CCNN). The details about the working of CCNN is available in our 
		previous work \citep{mace2}. 

\section{Observations and Data Analysis}
\label{sec:analysis}
		Observations of Crab Nebula were carried out using MACE 
		at various zenith angles between November 2022 to February 2023. 
		We have evaluated the performance of the telescope at zenith angles 
		ranging from 10$^{\circ}$ - $35^{\circ}$. The data were 
		collected in the ON/OFF mode for $\sim$ 15/ 12 hours 
		respectively. 
		The OFF sky region, 
		devoid of any known $\gamma$-ray source, was observed within the 
		same zenith range as that of ON source to collect the OFF source 
		data. After applying the Good Time Interval (GTI) (refer Section $~\ref{sec:dqm}$), we obtained 
		a total of 12.57 hours and 8.32 hours of ON and OFF source data respectively.

\subsection{Data Quality Monitoring}
\label{sec:dqm}
		The MACE console serves as a central control unit where the 
		system configuration parameters and operation sequences are 
		selected. Central Camera Control (CCC) collects the telemetry 
		data from all the CIMs and sends it to the Property Server 
		(PS) for data storage. Two types of telemetry parameters are 
		recorded based on their impact on the system performance: 
		the "Critical Parameters" like Anode Current (AC), Single 
		Channel Rate (SCR), System PCR (Promp Coincidence Rate) and 
		Module CCR (Chance Coincidence Rate), logged every second, and the 
		"Routine Parameters" like High Voltage (HV), Discriminator Threshold 
		(D$_{th}$), power supply Status and Temperature, logged every minute.
		In-house developed ROOT-based software is used to analyse the 
		Critical and the Routine telemetry data to carry out the data 
		quality checks for proper hardware functioning of MACE 
		camera. Analysis of telemetry data ensure that the hardware 
		is operating within the acceptable ranges. 

		To prepare the science data for the event analysis, a utility 
		called GTI was developed. This utility checks the variation 
		of the critical and the routine telemetry parameters and 
		selects time intervals where telemetry parameters remains 
		within the acceptable range. This utility is implemented 
		in three stages: Global GTI, which deals with parameters 
		for the entire camera, module GTI, which deals with 
		module-based parameters and channel GTI, which deals 
		with individual channel parameters. GTI on the telemetry 
		data generates good time intervals for the Critical and 
		the Routine parameters, combining them to form a single GTI 
		for a particular run. 

\subsection{Calibration \& Trigger}
\label{sec:cal}
		The event data (ON data) contains low-gain and high-gain charges from 
		all 1088 PMTs. High-gain channels utilize an additional amplifier 
		(with a gain of 10) compared to that of low-gain channels to 
		accommodate even the weaker Cherenkov pulses. The charges are calculated 
		from the 24 samples from the Domino Ring Sampler (DRS) version 4 \citep{Ritt2010}
		operating at rate of 1 GS/s. To reduce the data 
		volume, the charges are calculated in the camera itself. Pulse profiles 
		of 31 ns are stored only for the channels above a particular threshold. 
		In addition to the charge and profile, the dataset includes hardware 
		information and flags to ensure data reliability. 
		The Cherenkov, sky, and calibration data along with the timing information are stored in the same file 
		with corresponding flags. 
		The trigger rate of MACE follows zenith angle
		dependence of the form $\mathrm R(\theta) \sim \textrm R_{0}(\theta)\textrm cos^{m}(\theta)$,
		where $\mathrm \theta$	is the zenith angle of observation and $\mathrm m$ = 0.41.
		Observations with rates deviating more than 10\% with respect to 
		a value estimated from above relation were discarded
		from the dataset. We also discarded data affected by technical
		problems or poor sky conditions. The sky conditions are assessed
		based on data from the on-site sky monitoring system as well as
		the fluctuations in single channel rates. 
		The MACE employs an altitude-azimuth mount with 
		a 27 m diameter circular track that can provide the 
		tracking accuracy of better than 1 arc-minute under wind speeds of up to 30\, 
		km $\textrm h^{-1}$. 
		The pointing of the telescope is verified for each
		source by observing an optical star having similar declination as
		that of $\gamma$-ray source. The telescope also	incorporates an
		active mirror alignment control system to correct for any deformations
		due to gravity. The exact location of the star was compared with the reconstructed
		positions of the star in the camera after the pointing run. The
		final telescope mis-pointing was corrected offline during the pre
		and post-transit phase of Crab Nebula. The pointing accuracy was
		measured to be better than 1 arc-minute. 
		The MACE records 1000 calibration and 1000 sky pedestal events after every 300 
		seconds of collecting Cherenkov data. The LED calibration events are 
		used to estimate relative gain factors of channels and to 'flat-field' 
		images. The mean value and variance of the night sky background for 
		each channel are estimated from sky pedestal events. After calibration, 
		single bad channels (a specific channel that is malfunctioning or inactive. 
		This can be due to various issues such as a dead pixel, high or low anode 
		current, or other related problems.) that are part of the images are assigned values 
		obtained from the interpolation of neighboring channels.
 





\subsection{Image cleaning and reconstruction}

		The night sky background (NSB) plays a significant role in modulating the 
		performance of MACE. This background is commonly referred to as the 
		light of the night sky (LONS). During the moonless nights, the night sky background 
		consists of the diffuse light contributed by the ambient sources such as the surroundings, 
		starlight, and zodiacal light. The measurement of the LONS at the nearby 
		Himalayan Chandra Telescope (HCT) has been carried out \citep{Stalin}. The 
		intensity of the NSB in the wavelength range 240-650 nm is estimated to be 
		$\sim 2.6 \times 10^{12}$ photons m$^{-2} \textrm s^{-1} \textrm sr^{-1}$ \citep{mace1}. 
		To improve the quality of the Cherenkov light triggered images, 
		a two-stage image cleaning procedure is 
		employed using the standard cleaning method \citep{cleaning1996}. In the 
		first stage, a core set of image pixels are identified where the total charge 
		within the image exceeds a given threshold. The second stage involves 
		identifying the boundary pixels where total charge is more than a 
		given threshold. Standard tail-cut cleaning is applied to every 
		image where each pixel with Signal to Noise ratio (S/N) of 
		$\geq$ 6.5 are retained as picture pixels while the boundary pixels are selected 
		if they have S/N ratio of $\geq$ 2.65 and are adjacent to atleast one 
		picture pixel. The isolated pixels are retained only if their S/N ratio is 
		$\geq$ 10.0. while the remaining pixels are excluded from further analysis.

\begin{figure}
\begin{center}
\includegraphics[width=8.8cm,height=5.2cm,angle=0,clip]{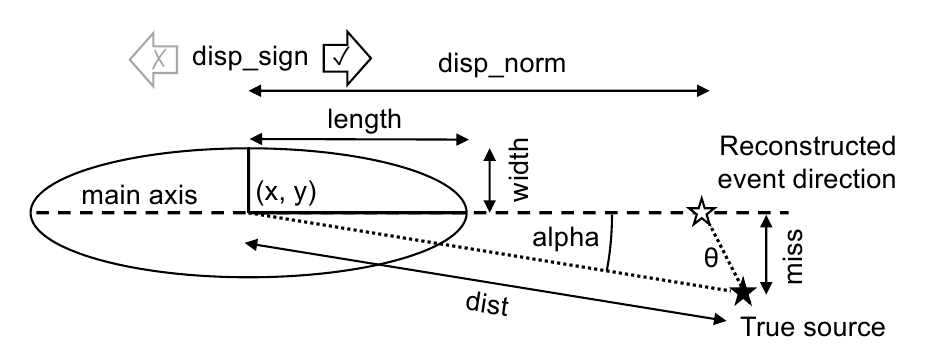}
	\caption{\label{fig:hpar} Illustration of few Hillas parameters of a typical Cherenkov image \citep{CTA2023}.}
\end{center}
\end{figure}

		Once the images are cleaned they are characterised by calculating
		the Hillas parameters \citep{Hillas}. These parameters are based on the zeroth, first
		and second moments of images weighed by charge content of the pixels.
		Some of the Hillas parameters used in this work as 
		inputs to the random forest models for event classification, direction 
		reconstruction and energy prediction are depicted in the Figure \ref{fig:hpar}.
		Additionally, we have used Leakage parameter (ratio between 
		the light content in the camera two outer most pixels to the total light 
		content) to quantify the extent of image truncation.

\begin{figure}
\begin{center}
\includegraphics[width=9.7cm,height=5.5cm,angle=0,clip]{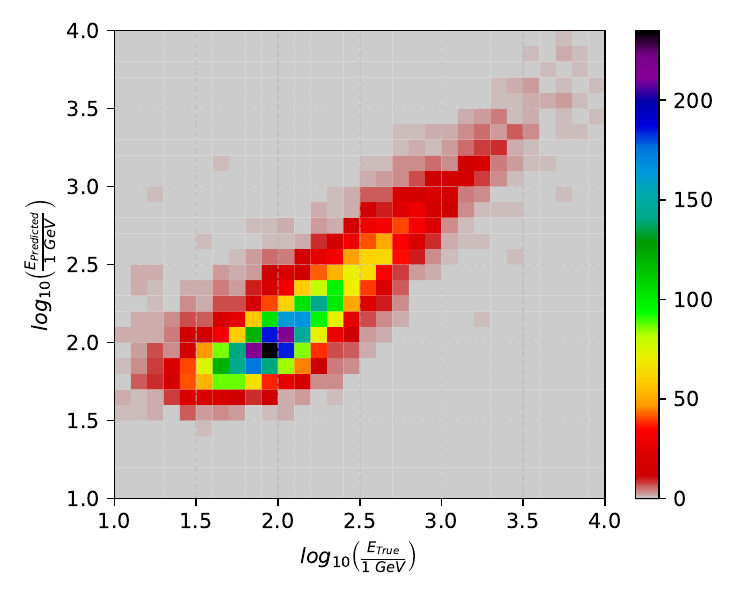}
	\caption{\label{fig:estE} Distribution of simulated $\gamma$ ray events 
	in True versus the estimated energy using the Random Forest method. 
	The color of the bin represents the number of $\gamma$-ray events.}
\label{Figure8}
\end{center}
\end{figure}

\subsection{Event Classification}
\label{sec:classification}
		The $\gamma$ and hadron events are distinguished by carefully analyzing 
		the spatial distribution of Cherenkov photons on the image plane of the 
		camera. Only a very small fraction ($\sim$ $10^{-5}$) of $\gamma$-ray 
		events constitute the signal compared to the background generated by 
		cosmic rays. Furthermore, the segregation of $\gamma$-ray events from the background 
		is complicated by the NSB triggered images, which can obscure the signal of 
		interest. We use multivariate techniques for the discrimination of $\gamma$-rays 
		and hadrons employing multiple variables to analyze the data and identify patterns 
		that distinguish $\gamma$-rays from hadrons. The Random Forest method (RFM) 
		\citep{rf2001} is employed for $\gamma$-hadron classification. The events 
		are classified using a single classification parameter, known as the {\it {hadronness}}. 
		The Hillas parameters like Length, Width, Distance, Size, Frac2 
		(ratio of the sum of two highest pixel signal to the sum of all the pixel signal), and Leakage 
		are used for training the Random Forest classifier employing scikit-learn \citep{scikit-learn}. 
		The values of {\it {hadronness}} range between 0 and 1 with higher 
		indicating that the classified event is more like a hadron event. 
		The value of {\it {hadronness}} is chosen to maximize the significance 
		of signal detection. In this work, we have applied a cut value of Hadronness 
		$\leq$ 0.14 to categorize the image as a $\gamma$-ray-like event.

\subsection{Energy reconstruction}
		The energy of $\gamma$-rays is reconstructed using the regressive RFM. 
		The RFM was trained using standard Hillas parameters (Size, Distance, 
		zenith, Leakage) and a simulated dataset consisting of 160,000 $\gamma$-rays, 
		as described in the simulation section. Figure ~\ref{fig:estE} displays the 
		estimated energy as a function of the true energy illustrating that the reconstructed 
		$\gamma$-ray energy closely aligns with the true $\gamma$-ray energies.

\begin{figure}
\includegraphics[width=4.4cm,height=6.0cm,angle=0,clip]{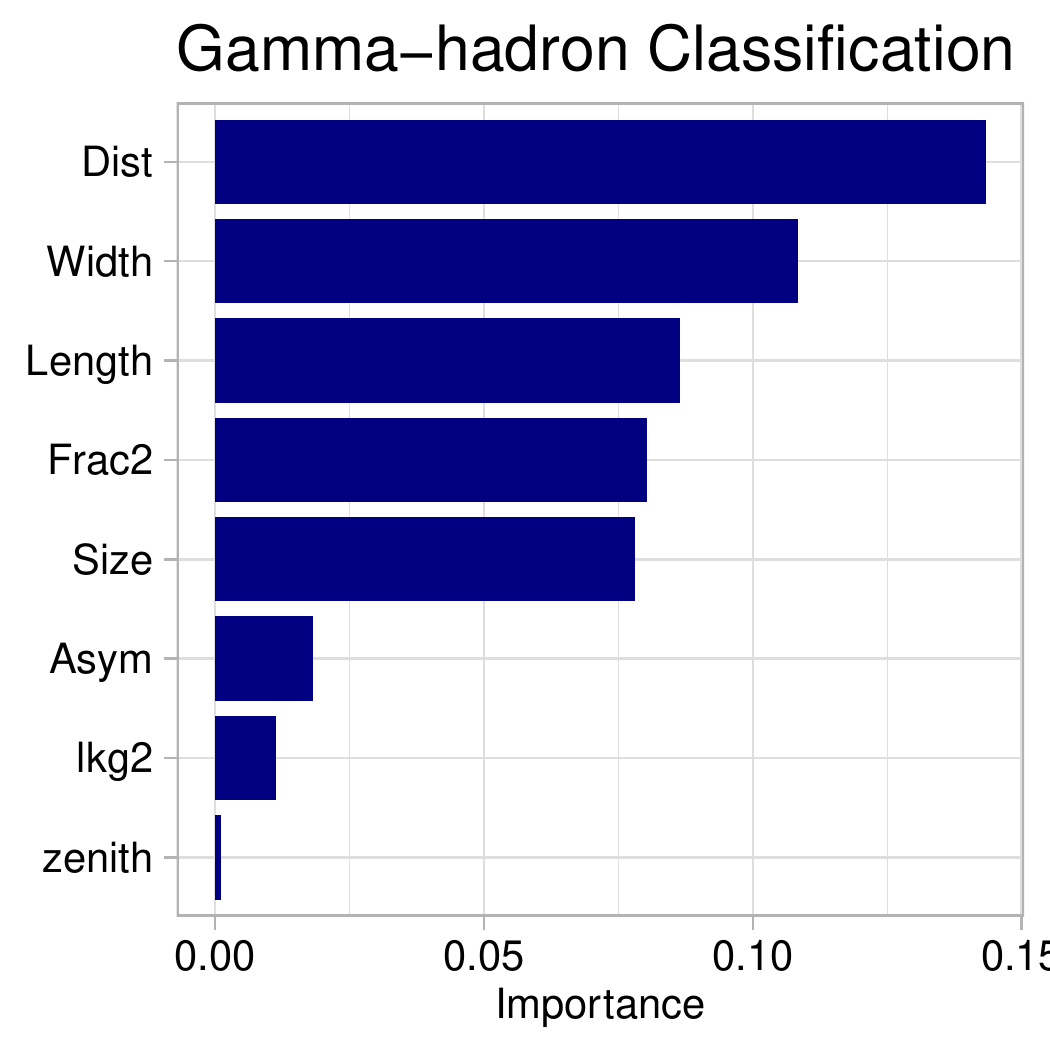}\includegraphics[width=4.4cm,height=6.0cm,angle=0,clip]{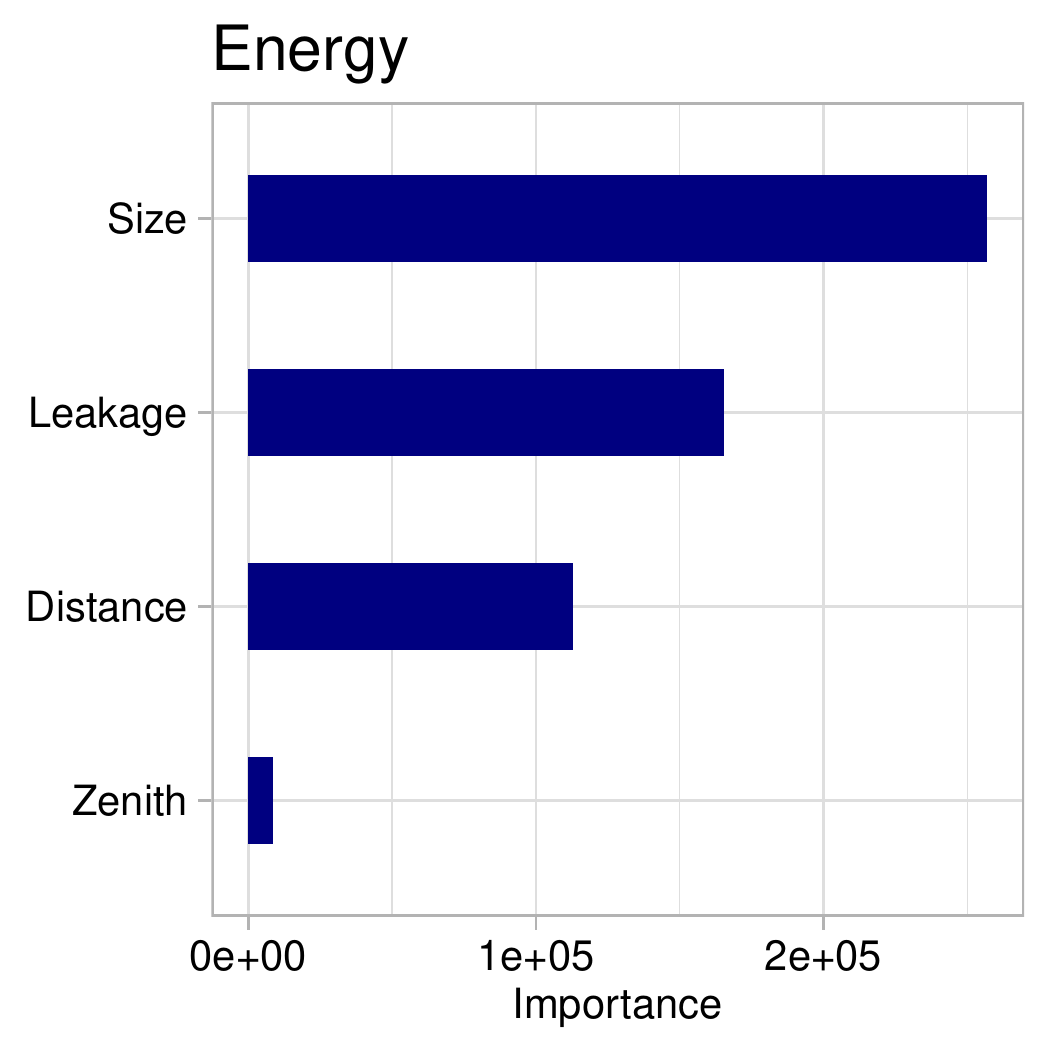}
	\caption{\label{fig:var} Relative importance of various parameters for gamma-hadron classification (Left panel) and energy reconstruction (Right panel).}
	\end{figure}

		With the RFM, we have estimated the relative importance \citep{varimp} of all the 
		training parameters (features) used in classification and the 
		energy estimation employing the simulated gamma and proton events. 
		The relative importance is determined by calculating the 
		mean decrease in accuracy, which measures the model performance 
		without that specific variable. A higher value indicates a 
		greater importance of that variable in prediction. The left panel of Figure~\ref{fig:var} 
		illustrates the relative importance of various parameters in classification. 
		The relative importance of various 
		parameters in energy estimation is shown in the right panel of 
		Figure~\ref{fig:var}. 

\section{Validation of the MC simulation by comparisons with observation data}
\label{sec:comparison}
\begin{figure}[h!]
\begin{center}
\includegraphics[width=8.0cm,height=5.9cm,angle=0,clip]{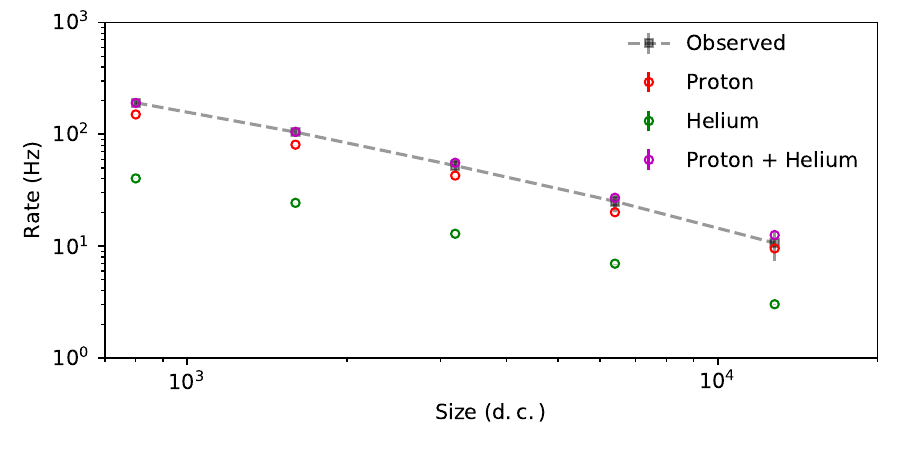}
	\caption{\label{fig:rate} Integral rate of background events above a given size
	for data (dashed line) and MC. The contribution of proton in the total rate is 
	shown with red open circles, green open circles show contribution of Helium 
	while magenta open circles show total simulated rates.}
\end{center}
\end{figure}

The MC chain was validated at 3 stages of analysis. This includes 
comparison of experimental event rates with simulated background rates and
comparison of experimental Hillas parameters distributions of background
and excess data with simulated background and $\gamma$-ray data. 

\subsection{Rate Comparison}
As a first step of validation of MC against the data,
we compare the integral event rates above a given size for the data and
the MC samples. The event rates of an IACT are dominated by proton and
Helium induced air showers. While the protons contribute nearly 80\% of
the total collected shower images, Helium induced showers
produce rest of the events. Figure \ref{fig:rate} shows the integral rates
due to Protons (red open circles), Helium (green open circles) and their total
(magenta open circles) as estimated by MC simulations. The grey dashed line 
and corresponding open squares show the variation of observed integral rates
as a function of size threshold. We see that, there is a good agreement between 
simulations and data. We mark here that, although calibrated values of PMT gain,
optical PSF and other electronics parameters have been incorporated in the 
simulations, uncertainties due to variations in atmospheric and sky conditions,
degradation of mirrors or their dust deposition can not be measured. Hence an
additional parameter of optical efficiency, also termed as telescope gain, is
introduced in the simulations for fine tuning. We find that the integral
rates estimated by MC simulations match well with the OFF data rates at the
optical efficiency value of 1.05.

\subsection{Hillas Parameters : Data vs MC}

\begin{figure}[h!]
\begin{center}
\includegraphics[width=0.49\textwidth,angle=0,clip]{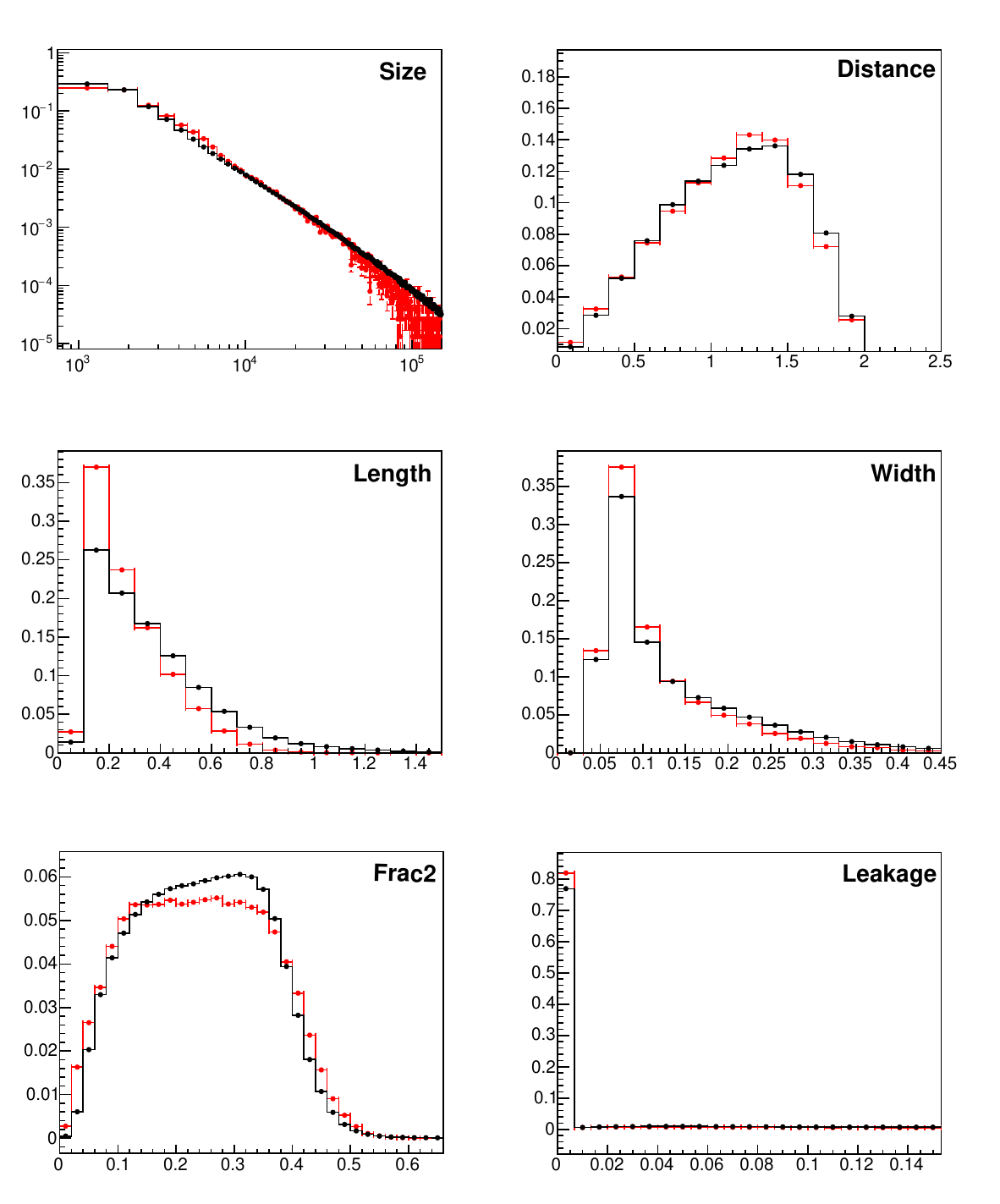}
	\caption{\label{fig:valp} Distribution of various Hillas parameters for the 
	 proton events. The black and red lines show the simulation and the
	observed data points respectively for Size, Distance, Length, Width, Frac2 and Leakage.}
\end{center}
\end{figure}
\begin{figure}[h!]
\begin{center}
\includegraphics[width=0.39\textwidth,angle=0,clip]{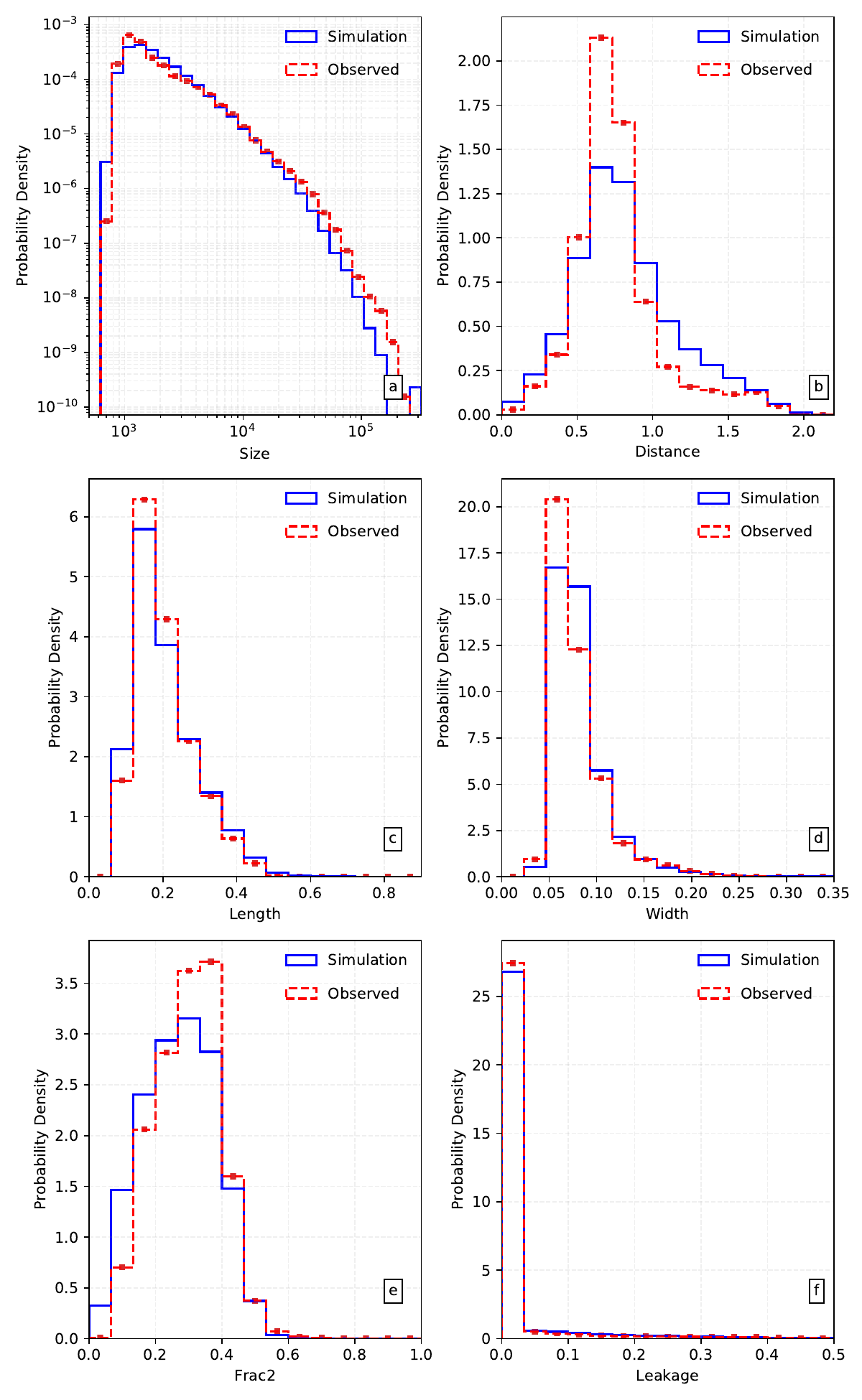}
	\caption{\label{fig:valg} Distribution of various Hillas parameters for the 
	$\gamma$-ray excess events extracted by using the RFM and Monte Carlo simulated 
	$\gamma$-rays. The solid lines (blue) and the dashed line (red) show the simulation and the
	observed data points respectively for Size, Distance, Length, Width, Frac2 and Leakage.}
\end{center}
\end{figure}
		In order to validate the performance of MACE, a comparison 
		of Hillas parameters for background events and the predicted 
		$\gamma$-ray-like events observed by MACE telescope Crab$_{OFF/ON}$ observations 
		and those obtained from MC simulations is carried out. A 
		representative comparison of the various image parameters like 
		Length, Width, Distance, Size, Frac2 and Leakage is shown in 
		Figure \ref{fig:valp}, \ref{fig:valg}.
		We find acceptable agreement between the simulated and the observed
		parameters. However, there is a visible deviation between simulated and 
		observed distributions of Size and Distance parameters at higher values. 
\begin{figure}
\begin{center}
\includegraphics[width=8.2cm,height=5.3cm,angle=0,clip]{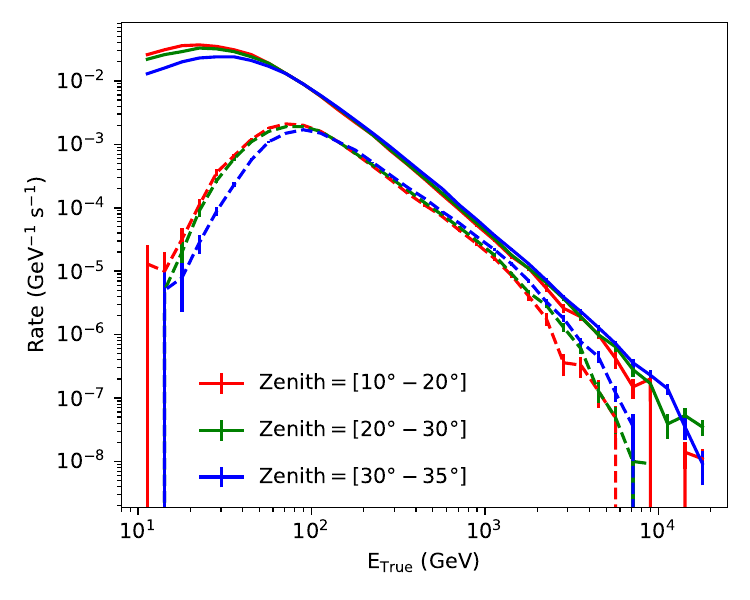}
	\caption{\label{fig:eth}Energy threshold of MACE: The differential 
	$\gamma$-ray rate as a function of the true energy. Dashed line curves 
	show $\gamma$-ray differential rate after applying the analysis cuts
	in the zenith angle range 10$^\circ$-20$^\circ$, 20$^\circ$-30$^\circ$
	and 30$^\circ$-35$^\circ$, while solid lines show rate curves for
	triggered events in the same zenith angle ranges. The energy
	threshold shifts from $\sim$ 30 GeV to $\sim$ 80 GeV after applying 
	the analysis cuts.}
\end{center}
\end{figure}

\section{MACE telescope performance}
\label{sec:perform}
		The performance evaluation of MACE utilized the MC simulations 
		and employed four metrics: energy threshold, energy resolution, angular 
		resolution, and integral sensitivity. The evaluation was carried out 
		using the Random Forest multivariate machine learning method.
		We mark that we assume power law spectral shape for
		the Crab nebula with index of 2.6, during the estimation of energy
		threshold and integral sensitivity.

\subsection{Energy Threshold}
		The threshold energy of a telescope is determined by 
		constructing a differential rate plot of $\gamma$-rays
		utilizing the Monte Carlo simulations as a function of energy. The 
		energy at which this rate plot peaks, signifies 
		the energy threshold ($E_{th}$) of the telescope. Figure ~\ref{fig:eth} 
		displays the trigger rate plot, modelled by a power law, as a function of true $\gamma$-ray 
		energy before and after the analysis cuts in the 
		zenith angle range 10$^\circ$-20$^\circ$, 20$^\circ$-30$^\circ$ and 
		30$^\circ$-35$^\circ$. The $\gamma$-ray rate plot shows a broad peak, 
		indicating an energy threshold of $\sim$ 80 GeV. 
		Because the peak of $\gamma$-ray rate is relatively broad and 
		extends significantly towards the lower energies, it remains possible 
		to have events below the energy threshold of the telescope. However, 
		below 80 GeV, there is a reduction in the effective area of the telescope 
		(Figure ~\ref{fig:effarea}), making it challenging to 
		analyze faint sources within this energy range.
\begin{figure}
\begin{center}
\includegraphics[width=8.2cm,height=5.8cm,angle=0,clip]{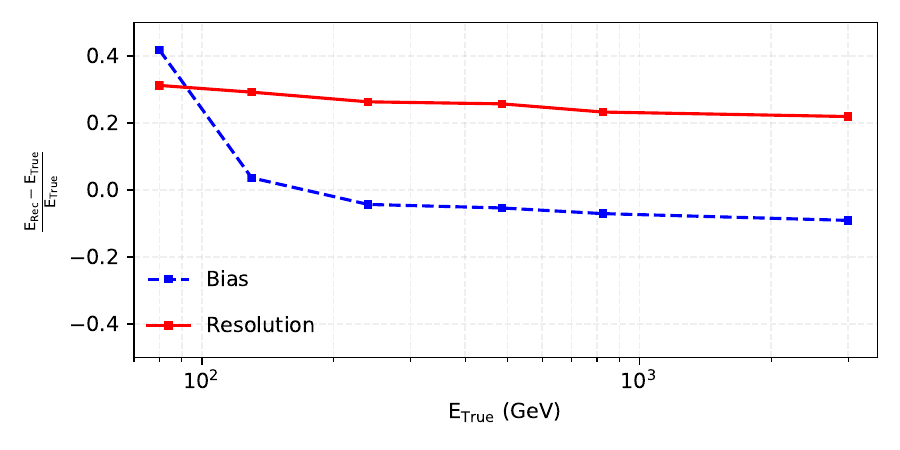}
	\caption{\label{fig:ER}Energy resolution (solid line) and bias (dashed line) of MACE
	as a function of the true energy of Monte Carlo averaged over zenith
	angle range of 10$^\circ$ - 35$^\circ$.}
\end{center}
\end{figure}
\subsection{Energy Resolution}
            The inherent fluctuations in the extensive air shower generation process, 
			as well as the effects of the atmosphere and Earth's magnetic field, 
			have an impact on an IACT's capability to accurately reconstruct 
			the energy of primary $\gamma$-rays. These disturbances can 
			modify the physical characteristics of the shower, potentially leading 
			to a spread in the signals, and consequently introducing a bias in 
			the reconstructed energy of the $\gamma$-rays. The energy resolution 
			is defined as the standard deviation of the 1-D Gaussian distribution fitted 
			to the distribution of the relative fractional difference between the true and 
			estimated energy. It is expressed as a function of $\gamma$-ray energy, 
			($\frac{(E_{est} - E_{true})}{E_{true}} = \frac{\delta E}{E_{true}}$). 
			To estimate the energy resolution, the true $\gamma$-ray energies are
			divided into energy bins. In each bin, the fractional 
			energy $\frac{\delta E}{E_{true}}$ is calculated and fitted with 
			a Gaussian function. The energy bias is defined as the mean of the
			distribution of fractional deviation of reconstructed energy
			relative to true energy. Figure \ref{fig:ER}
			shows the variation of the energy resolution (solid red curve) and
			energy bias (dashed blue line) as a function of true energy of 
			$\gamma$-ray events.The energy resolution is determined
			to be $\sim$ 34\% at the energy threshold ($\sim$80 GeV), which
			improves to $\sim$ 21\% above 1 TeV. The energy bias in the corresponding
			energy range changes from $\sim$ 42\% to $\sim$ -5\%. 

\begin{figure} 
\begin{center}
\includegraphics[width=8.2cm,height=5.9cm,angle=0,clip]{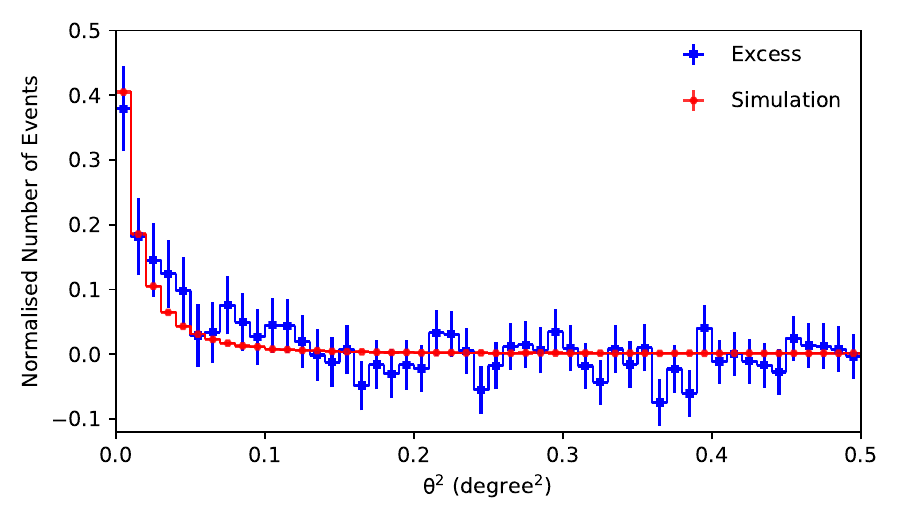}
	\caption{\label{fig:theta2} Comparison of background subtracted $\mathrm \theta^{2}$ distributions 
	for Crab nebula observations (blue line, filled square) and MC $\gamma$-ray
	events (red, filled circles).}
\end{center}
\end{figure}

\subsection{Point Spread Function (PSF)}
\label{AR}
		The $\gamma$-ray PSF refers to the distribution or spread of 
		reconstructed directions of $\gamma$-ray events from a point-like
		$\gamma$-ray source, as detected by the telescope. This distribution
		peaks at the true source direction, and the spread around the mean
		value arises due to statistical fluctuations and instrument effects. 
		The error in the reconstruction of the source position is determined by 
		the square of the angular distance between the nominal source position
		and the reconstructed source position in camera plane, denoted by $\mathrm \theta^2$. 

\begin{figure} 
\begin{center}
\includegraphics[width=8.0cm,height=5.9cm,angle=0,clip]{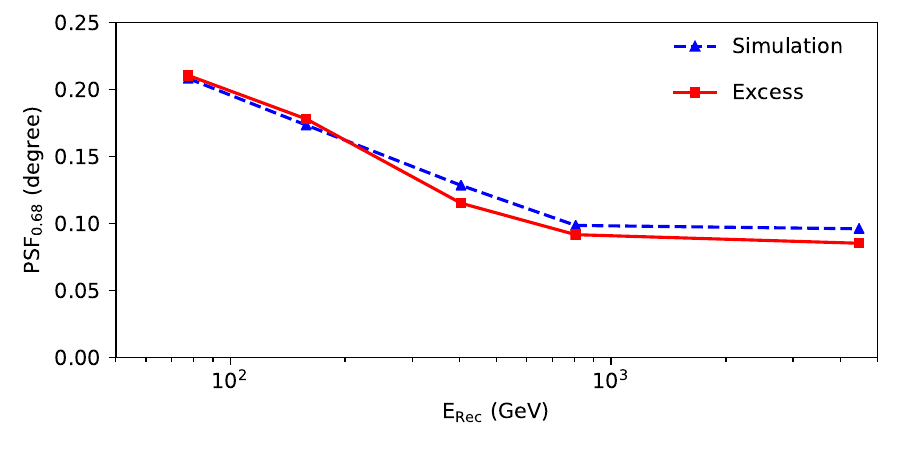}
	\caption{\label{fig:AR}The $\gamma$-ray PSF as a
	function of reconstructed energy from observations (solid line) and simulations
	(dashed line). It represents the 68\% containment radius for the distribution
	of reconstructed arrival directions around the nominal source location in 
	camera plane.}
\end{center}
\end{figure}
		\noindent We evaluate the PSF of MACE for point-like
		sources by estimating 68\% containment radius for the distribution
		of reconstructed positions around the nominal source location in
		camera. In case of observations the said quantity can be
		obtained by calculating the 68 percentile value of the $\mathrm \theta^{2}$ 
		distribution of excess events. Figure \ref{fig:theta2} shows the distribution
		of $\mathrm \theta^{2}$ in excess events obstained from Crab nebula observations. 
		The excess events are calculated by subtracting the $\mathrm \theta^{2}$ distribution
		during OFF runs from $\mathrm \theta^{2}$ distribution obtained during ON run.
		The same figure shows the distribution of $\mathrm \theta^{2}$ for simulated
		$\gamma$-ray events. The hadron score cut of $\mathrm h <= 0.14$ 
		is applied in addition to quality cuts when finding the $\mathrm \theta^{2}$
		distributions of data excess as well as simulations. The figure shows 
		a good agreement between the simulated and observed extension of point
		source.
		
		\noindent Figure \ref{fig:AR} shows the variation of
		$\gamma$-ray PSF as a function of reconstructed energy $\gamma$-ray
		events, obtained from data and simulations respectively.
		The two values are consistent with each other, with PSF improving
		from $0.20^{\circ}$ near energy threshold to $0.08^{\circ}$ above
		energy of 1 TeV.
		
\subsection{Integral flux sensitivity}
\begin{figure} 
\begin{center}
\includegraphics[width=8.05cm,height=5.2cm,angle=0,clip]{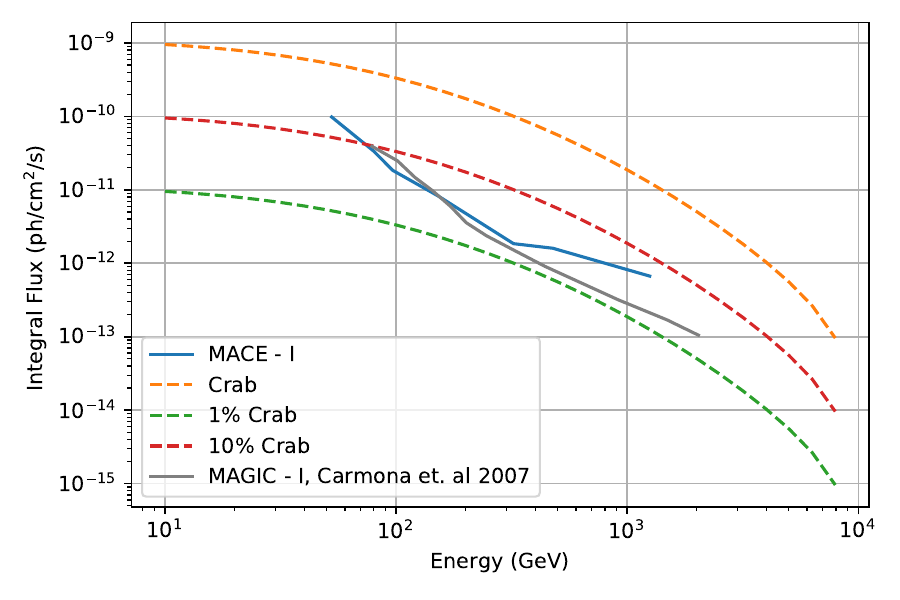}
	\caption{\label{fig:IntS} The Integral Sensitivity (the integrated flux of a source above given energy) 
	of MACE, assuming log-parabola spectral shape as a function of true $\gamma$-ray energy in the zenith angle 
	range $10^\circ-35^\circ$, subject to a condition that the minimum number 
	of excess $\gamma$-ray like events $\textrm N_{excess} > 10$, and ($\textrm N_{excess} > \textrm 0.05 \textrm N_{bkg}$). 
	The fractions of the integral Crab Nebula flux are also plotted (1\% and 10\%) along 
	with the integral flux from MAGIC-I \citep{magic1} is also shown.}
\end{center}
\end{figure}

		The Integral sensitivity determines the minimum time required 
		for significant detection of a $\gamma$-ray source. The minimum 
		integral photon flux emitted by a Crab Nebula like source 
		detectable above a certain energy threshold by an IACT with a 
		5$\sigma$ significance in 50 hours of observation, is known as 
		the \emph{Integral flux Sensitivity} of a telescope, subject 
		to the condition that the minimum number of excess $\gamma$-ray 
		like events must be at least 10 ($\textrm N_{excess} > 10$), and 
		($\textrm N_{excess}$ $>$ 0.05$\textrm N_{bkg}$). Mathematically, the telescope 
		sensitivity is defined as the flux of a source with 
		$N_{excess}$ / $\sqrt{N_{bkg}}$ = 5 after 50 hours of effective 
		observation time. The estimated integral sensitivity of MACE 
		for a Crab Nebula-like source in the zenith angle range 
		$10^\circ-35^\circ$ above a given energy threshold ($E_{th}$) is 
		given in Figure~\ref{fig:IntS}. Additionally, the integral 
		sensitivity of MACE is compared with that of MAGIC-I 
		telescope \citep{magic1} along with energy-dependent 1\% and 
		10\% Crab Nebula flux. MACE achieves the best sensitivity of 
		$\sim$ 1.75\% of the Crab Nebula flux at $\sim$ 320 GeV. The 
		integral sensitivity of MACE near its energy threshold 
	($E_{\gamma} \sim$ 80 GeV) is estimated to be $\sim$ 3.46\% 
		of the Crab Nebula flux. It is to be noted that the 
		sensitivity is obtained from ON-observation time only.
		The quoted MACE and MAGIC-I sensitivity is estimated assuming
		power law spectral shape for Crab nebula as reported in \citep{magic1}.
		On the other hand, the integral flux curves shown for various 
		fractions of Crab flux are calculated with log-parabola spectral 
		shape as reported in \citep{Albert2008}. It is evident from Figure
		\ref{fig:IntS} that MACE sensitivity is $\sim$ 10.0\% if we
		assume log-parabola spectral shape for Crab nebula.

\section{Results and conclusion}
\label{sec:results}
		In this section, we present the results obtained by the data analysis of Crab Nebula 
		observations using MACE telescope and calculated the light curve, 
		spectral energy distribution and the flux sensitivity.
		 
		\subsection{Experimental flux sensitivity}
\begin{figure}
\begin{center}
\includegraphics[width=8.0cm,height=5.4cm,angle=0,clip]{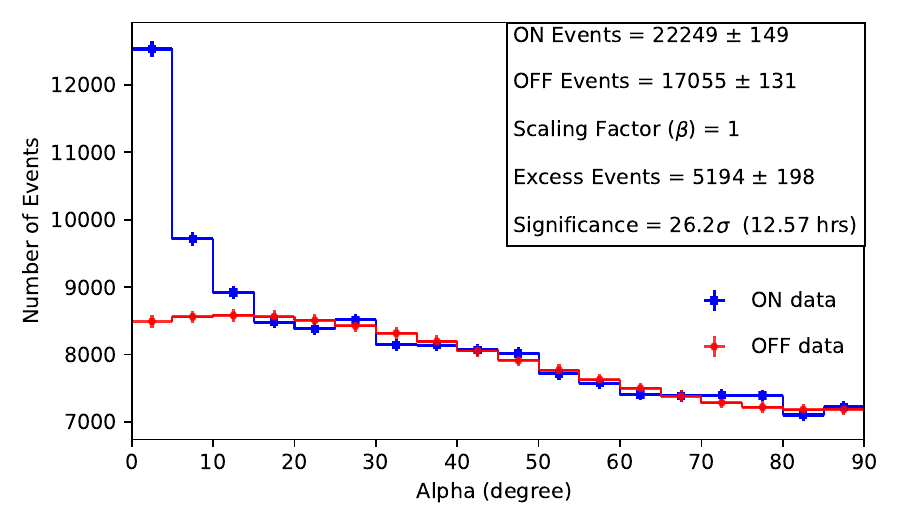}
		\caption{\label{fig:alp} Distribution of the image orientation parameter 
		$\alpha$ for $\gamma$-ray events (blue: ON data; red: OFF data) as predicted 
		using the RFM. The $\gamma$-ray signal is extracted from the $\alpha$ 
		bin of 10$^\circ$ whereas the background region is considered within 
		$25^\circ$ $\leq$ $\alpha$ $\leq$ $85^\circ$. The OFF data 
		is normalized between 25 and 85 degrees alpha with respect to the ON 
		data. Subsequently, the ON - OFF information from each bin is
		utilized to determine the signal.} 
\end{center}
\end{figure}

		For a single imaging Cherenkov telescope, source dependent analysis
		based on the nominal source position in the camera is more sensitive.
		Since the IACTs operate in the source tracking mode, the orientation parameter
		$\alpha$ is very small for the excess $\gamma$-ray events, indicating that 
		most of the $\gamma$-ray events arrive from the same direction. 
		Consequently, this parameter plays a crucial role in signal 
		extraction. The $\gamma$-ray-like events are selected using 
		the $\alpha$ after $\gamma$-hadron classification. Figure~\ref{fig:alp} 
		displays the distribution of events as a function 
		of this parameter after application of gamma-domain and quality cuts. We note that,
		the average gamma acceptance for the gamma-domain cuts is estimated to be 41.2\% while 
		proton rejection is found to be 99.03\%. The signal is extracted from the $\alpha$ bin 
		of 10$^\circ$ whereas the background region is considered within 
		$25^\circ$ $\leq$ $\alpha$ $\leq$ $85^\circ$. We have detected an 
		excess of 5194 $\pm$ 198 $\gamma$-ray-like events with a 
		statistical significance of 26.2 $\sigma$ in 12.57 hours. These values
		of significance and time of observation translate to an integral sensitivity
		of $\sim$ 9.6\% which are consistent with value projected by simulations. 

\subsection{Differential Energy Spectrum of Crab Nebula}
		The obtained light curve and the differential energy spectrum are
		the two important metrics to establish the proper functioning of the
		telescope.
	\noindent
		The performance of MACE was further evaluated by estimating 
		the differential energy spectrum. 
Two factors must be taken into account while determining the differential
		energy spectrum of the VHE gamma-rays from their observed energy distribution,
		namely, the finite resolution and finite bias of the energy reconstruction
		process in imaging Cherenkov telescopes. These factors cause the ‘migration’
		of VHE gamma-ray events from their true energy bins to the reconstructed
		energy bins. The migration of events from true energy bins to nearby reconstructed
		energy bins is higher at low energies, while it decreases with increasing energy.
		These effects result in observation of distorted frequency distribution in
		reconstructed energy as compared to their true energy frequency distribution.
		The number of observed gamma ray events in reconstructed energy bins are given by 
		\begin{figure}
		\begin{center}
		\includegraphics[width=8.4cm,height=5.2cm,angle=0,clip]{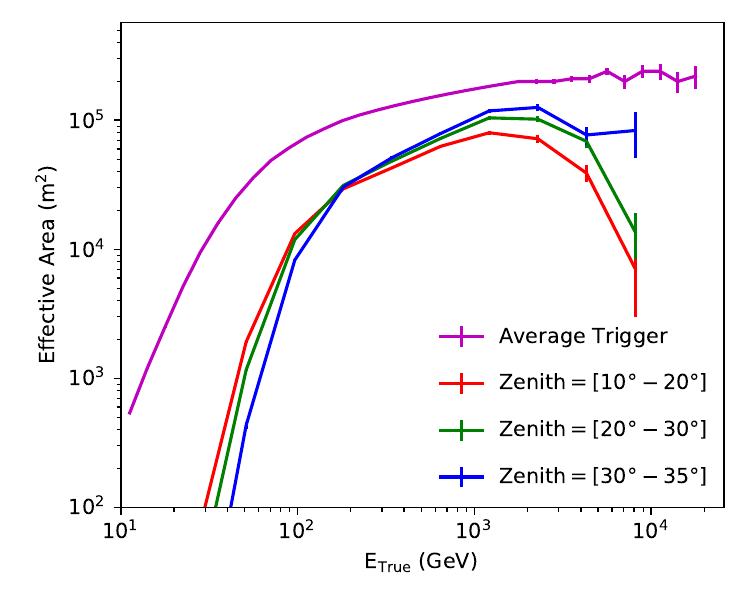}
		\caption{\label{fig:effarea} The effective collection area of MACE at 
			the trigger level (magenta color) and after applying the analysis cuts (red, green, blue color) for 
		the zenith angle range $10^{\circ}$ - $20^{\circ}$, $20^{\circ}$ - $30^{\circ}$, 
			$30^{\circ}$ - $35^{\circ}$ respectively.}
		\end{center}
		\end{figure}
		 \begin{equation}
		Y(E_{i}) = \sum \limits_{j=1,k=1}^{j=n_{a},k={N}} A_{j}^{k}(E_{j}) M_{ij}^{k} (E_{j},E_{i}) S(E_{j})  
		\label{eq:mig}
		\end{equation}
		Where, $Y(E_{i})$ are the total observed events in ith reconstructed energy bin,
		$A_{j}^{k}(E_{j})$ is the effective area of the telescope for jth
		true energy bin during the $k^{th}$ spell of observation,
		$M_{ij}^{k}(E_{j} → E_{i})$ represents the element of migration matrix
		for $k^{th}$ spell, which gives probability that an event in ith bin
		of reconstructed energy came from $j^{th}$ bin of true energy, $S(E_{j})$
		are the number of events in the $j^{th}$ bin of true energy, $n_{a}$ is
		number of bins in true energy, and $N$ is the total number of observation
		spells into which data is divided. We have used the method of forward unfolding
		\citep{forwardfolding}, to estimate differential spectrum of gamma-rays from the
		observed spectrum in reconstructed energy. The method uses the parametric form
		$S(E_{j} ; \hat{q})$ for source spectrum where $q$ represents a set of parameters
		of a spectral shape to be fitted. A $\chi^{2}$ statistic is then minimized
		to find the best fit values of parameters $\hat{q}$, where $\chi^{2}$ is given by,
        \begin{equation}
			\chi^{2} = \sum \limits_{i=1}^{i=n_{b}} \frac{(O(E_{i}) - \hat{Y}(E_{i}) )^{2}}{\sigma_{i}^{2}}
		\label{eq:chi2}
		\end{equation}	
		Where, $O(E_{i})$ is the observed number of events in $i^{th}$ bin
		of reconstructed energy, $\hat{Y}(E_{i})$ is the estimated counts in
		$i^{th}$ bin of reconstructed energy based on equation \ref{eq:mig}
		and $\sigma_{i}^{2}$ are errors in observed events in $i^{th}$ bin
		of reconstructed energy. The effective area and migration matrices required for the forward
		unfolding of MACE telescope data were estimated from the $\gamma$-ray
		simulation	datasets.
\begin{figure}[h!]
\begin{center}
\includegraphics[width=8.1cm,height=5.2cm,angle=0,clip]{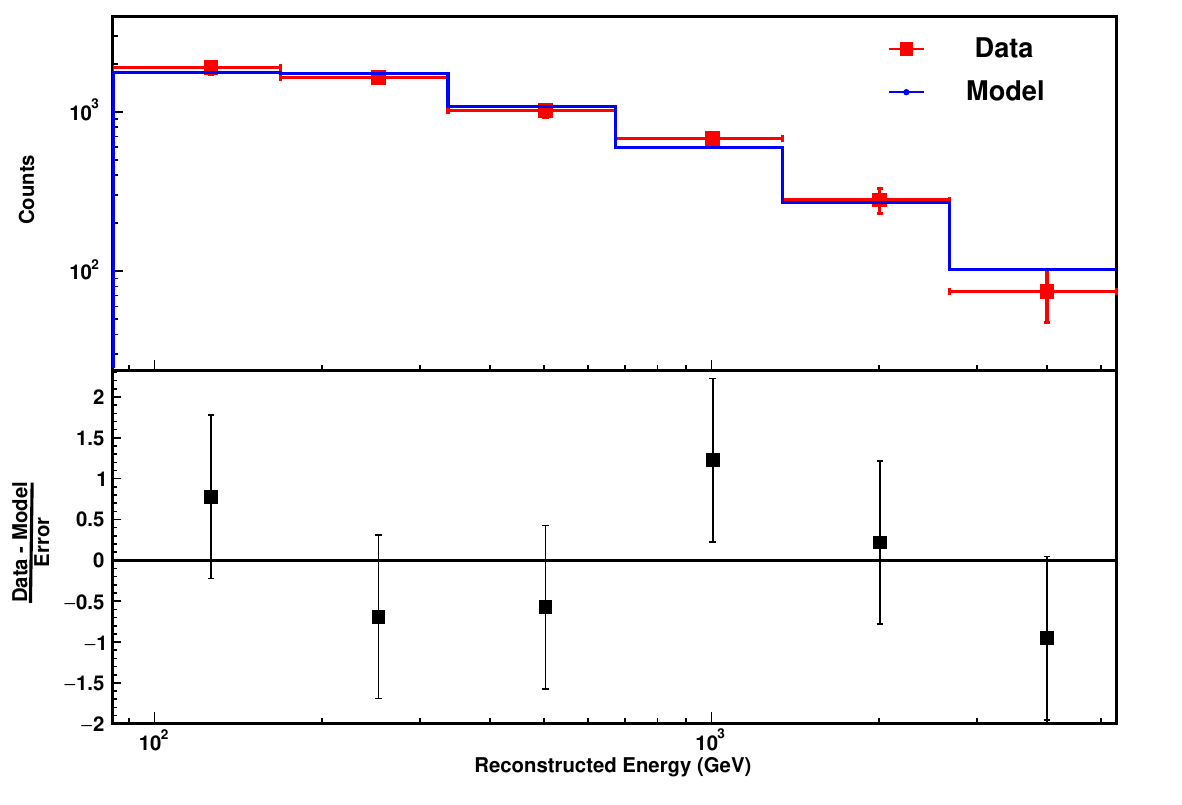}
\caption{\label{fig:spectrum_fit} Comparison of observed $\gamma$-ray excess
	counts with $\gamma$-ray counts estimated based on forward folded \citep{forwardfolding}
	log-parabola spectral model in different bins of reconstructed energy.}
\label{Figure6}
\end{center}
\end{figure}
\begin{figure}[h!]
\begin{center}
\includegraphics[width=8.7cm,height=5.5cm,angle=0,clip]{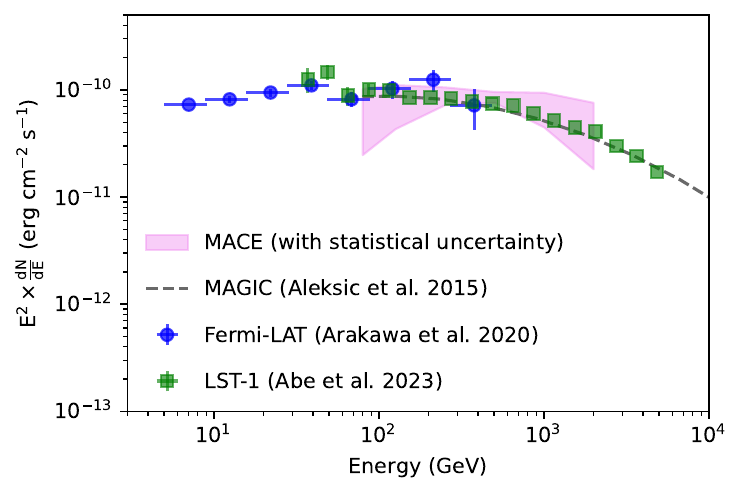}
\caption{\label{fig:sed}The Spectral energy distribution of the Crab Nebula obtained by MACE 
		(pink shaded region) and its comparison with that obtained by the Fermi-LAT, MAGIC, and LST 
		prototype \citep{CTA2023}.}
\label{Figure6}
\end{center}
\end{figure}
		 We divide the dataset into 3 zenith bin ranges
		of 10-20, 20-30 and 30-35 degrees respectively. Figure~\ref{fig:effarea} shows the
		effective area in these 3 zenith ranges as a function of energy,
		while average migration of events from true energy bins to reconstructed
		energy bins is depicted in Figure \ref{fig:estE}. 

		We use the log-parabola model of differential spectrum, defined as 
		\begin{equation}
			\frac{dN}{dE} = F_{0}(E/E_{0})^{(-\alpha – \beta log_{10}(E/E_{0}))}
		\label{eq:logparabola}
		\end{equation}
		to describe MACE observed energy distribution and compare it
		with spectrum reported by other groups. The value of $E_{0}$ is chosen
		to be 400 GeV, the energy at which the correlation between
		the flux normalization and spectral index is minimum. We fit the observed count spectrum in 
		the reconstructed energy range of $\mathrm 80 ~ GeV < E_{est} < 5 ~ TeV$. The best fit
		values for the parameters are: $F0 = (3.46 \pm 0.26) \times 10^{-10}$ $\textrm TeV^{-1} \textrm sec^{-1} \textrm cm^{-2}$ ;
		$\alpha = 2.09 \pm 0.06 $ and $\beta = 0.08 \pm 0.07$ with
		$\chi^{2}$ value of 4.2 for 3 degrees of freedom. Figure \ref{fig:spectrum_fit} shows 
		the comparison between observed excess and estimated $\gamma$-ray counts in different 
		bins of reconstructed energy values.
		The differential flux spectrum of Crab nebula obtained by MACE
		telescope shows reasonable agreement with measurements reported by other
		groups. The spectral slope measured by MACE telescope is little
		harder compared to the slope reported in literature \citep{MAGICCrab2012},
		however these values are still within the systematic and statistical
		errors of each other. It should also be noted that, the systematic
		uncertainties in MACE observations remain to be estimated.

		The spectrum of the Crab Nebula is also presented in the form of 
		Spectral Energy Distribution (SED) and compared with the SED obtained 
		by different telescopes operating in the GeV -TeV energy range, shown in 
		Figure \ref{fig:sed}. We find that
		the SED as measured by MACE is in reasonable agreement with SED measurements
		of other telescopes. We mark that , high error in the curvature parameter
		of the spectrum reflects in large error bars, specially at TeV 
		energies. 

\begin{figure}
\begin{center}
\includegraphics[width=8.1cm,height=5.9cm,angle=0,clip]{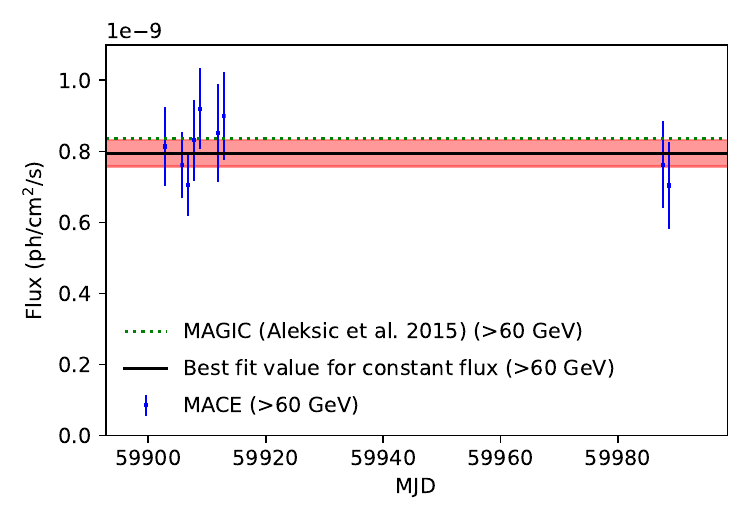}
		\caption{\label{fig:lcurve} Daily light curve and the statistical error band 
		in the measured mean $\gamma$-ray flux of the Crab Nebula for 
		energies above 60 GeV.} 
\end{center}
\end{figure}
\subsection{Lightcurve}
		The stability of the VHE $\gamma$-ray flux from the 
		Crab Nebula over the time scales of days and months was assessed by
		estimating the light curve of the Crab Nebula above 60 GeV, from
		MACE observation. Figure~\ref{fig:lcurve} displays the light 
		curve showing the daily flux measurements of Crab Nebula from 
		November 15, 2022, to February 23, 2023. The light curve 
		demonstrates the remarkable stability over both days, and 
		months, suggesting that the emission from Crab
		Nebula remains relatively constant throughout the observation.
		The small statistical fluctuations indicate the accuracy
		of measurements from MACE. To compare the daily fluxes of Crab
		Nebula using MACE and the MAGIC telescope, we estimated
		the integral flux from the MAGIC \citep{CrabMagic}
		above 60 GeV, modelled by the log-parabola function and using the
		best fit parameters. We obtained an integral flux of 
		$(8.35 \times 10^{-10}$\hspace{0.5 mm} ph $\textrm cm^{-2} \textrm s^{-1}$ 
		from the MAGIC telescope compared to a flux of (7.92 $\pm$ 0.36) 
		$\times 10^{-10}$ \hspace{0.5 mm} ph $\textrm cm^{-2} \textrm s^{-1}$
		as measured by MACE. 

\subsection{Conclusion}
\label{sec:concl}
        In this work, we report the results from Crab Nebula 
		observations by the newly commissioned state-of-the-art 
		IACT MACE at Hanle, Ladakh, India. 
		We carried out an extensive MC simulations using the 
		standard CORSIKA package to model the optical and 
		electronics response of MACE telescope. All the 
		analyses were performed using the in-house developed 
		MACE data Analysis Package (MAP). We carried out a comparison 
		of the Hillas parameters of the observed data against 
		those of MC simulations and found them to be in close 
		agreement. We have also investigated intra-night flux 
		variability and found that the emission from the source 
		remained stable, as anticipated. The integral flux 
		sensitivity of the telescope is $\sim$9.6\% 
		C.U. above 80 GeV over 50 hours of observations. The best 
		integral sensitivity is achieved for E$_{\gamma} \geq 320$ 
		GeV, which corresponds to $\sim1.75\%$ of the Crab Nebula flux. 
		The differential energy spectrum from the 
		Crab Nebula measured by MACE in the energy range 80 GeV 
		to 5 TeV is accurately described by a log-parabola function.  
		The spectrum aligns well with the spectrum obtained by the 
		MAGIC telescope within the bounds of systematic and statistical uncertainties. 
		The SED obtained using MACE telescope is in good 
		agreement with the previous measurements like MAGIC, LST-1 of 
		CTA and Fermi-LAT. An improved analysis method that incorporates 
		the arrival timing of the images will enhance our current 
		estimates. We anticipate an improvement towards the lower 
		energy end of the spectrum following a dedicated low-energy 
		analysis, which will be reported in a subsequent publication. 
		Furthermore, we would like to highlight that since 
		commissioning, all the technical components of the 
		MACE have performed as expected. 

\bibliographystyle{elsarticle/h}
\bibliography{mace}

\section*{Acknowledgements}
\noindent
The authors sincerely express their profound gratitude to Dr. A. K.
Mohanty, Chairman, Atomic Energy Commission, and Secretary, Department of
Atomic Energy (DAE), for his deep involvement, continuous support and
encouragement during the commissioning phase of the MACE telescope. We are
grateful to Shri Vivek Bhasin, Director, Bhabha Atomic Research Centre
(BARC) for his inspiring supervision and support in running the
observatory at Hanle. We gratefully acknowledge the valuable guidance,
time and efforts provided by Dr. S. M. Yusuf, Director, Physics Group/BARC
throughout the successful commissioning and smooth operation of the
telescope. We would like to acknowledge the unparalleled contributions made
by our colleagues from the Electronics Division, Control and
Instrumentation Division, Center for Design and Manufacture, and Computer
Division at BARC through their involvements in the indigenous development
of the various subsystems of MACE. We thank Dr. Stefan Ritt, Paul Scherrer
Institute, Switzerland for technical support in the utilization of Domino
Ring Sampler chips  in the MACE data acquisition system. We extend our
sincere gratitude to the two anonymous reviewers for their critical and
insightful comments and suggestions which have undeniably played a pivotal
role in the overall improvement of the manuscript.


\section*{Data Availability}
\noindent
		As an established norms of the field of ground based $\gamma$-ray 
		astronomy, regrettably, as of now, the raw data and calibration 
		files generated during various stages of MACE experiment are 
		presently unavailable for external access. The integrity of our 
		research has been maintained by safeguarding all the technical 
		records meticulously.









\label{lastpage}
\end{document}